\documentclass{ecai}

\usepackage{caption}
\usepackage{latexsym}
\usepackage{amsthm}
\usepackage{booktabs}
\usepackage{enumitem}
\usepackage{graphicx}
\usepackage{color}

\usepackage{amsmath}
\usepackage{amssymb}
\usepackage{booktabs,multirow}
\usepackage{mathtools}
\DeclarePairedDelimiter\ceil{\lceil}{\rceil}
\usepackage{wrapfig}
\usepackage{amsthm}
\usepackage{blkarray}

\usepackage{amsfonts}
\usepackage[qm]{qcircuit}
\usepackage{textcomp}
\usepackage{xcolor}
\usepackage{multirow}

\usepackage{newfloat}
\usepackage{listings}

\lstdefinelanguage{stripspddl}
{
  sensitive=false,
  morecomment=[l]{;}, 
  alsoletter={\#,:},   
  morekeywords={not, and, :action, :parameters, :precondition, :effect, :constants, :init, :goal, :types, :predicates, forall, when}
}

\DeclareMathOperator{\EO}{EO}
\DeclareMathOperator{\AMO}{AMO}
\DeclareMathOperator{\ALO}{ALO}
\DeclareMathOperator{\ctrlvar}{ctrl}
\DeclareMathOperator{\trgtvar}{trg}
\DeclareMathOperator{\cnotvar}{cnot}
\DeclareMathOperator{\matrixvar}{m}
\DeclareMathOperator{\colmatrixvar}{c}
\DeclareMathOperator{\forallrow}{R}
\DeclareMathOperator{\rowvar}{r}

\DeclareMathOperator{\cppairs}{CP} 

\DeclareMathOperator{\fullrankmatrix}{M}

\DeclareMathOperator{\numqubits}{n}
\DeclareMathOperator{\bin}{bin}

\newtheorem{definition}{Definition}

\newcommand{\BibTeX}{B\kern-.05em{\sc i\kern-.025em b}\kern-.08em\TeX}

\begin{document}

\begin{frontmatter}


\paperid{1524}


\title{Optimal Layout-Aware CNOT Circuit Synthesis with Qubit Permutation}


\author[A,B]{\fnms{Irfansha}~\snm{Shaik}}
\author[A]{\fnms{Jaco}~\snm{van de Pol}}

\address[A]{Department of Computer Science, Aarhus University, Denmark}
\address[B]{Kvantify ApS, DK-2300 Copenhagen S, Denmark}


\begin{abstract}
CNOT optimization plays a significant role in noise reduction for Quantum Circuits.
Several heuristic and exact approaches exist for CNOT optimization.
In this paper, we investigate more complicated variations of optimal synthesis by allowing qubit permutations and handling layout restrictions.
We encode such problems into Planning, SAT, and QBF.
We provide optimization for both CNOT gate count and circuit depth.
For experimental evaluation, we consider standard T-gate optimized benchmarks and optimize CNOT sub-circuits.
We show that allowing qubit permutations can further reduce up to 56\% in CNOT count and 46\% in circuit depth.
In the case of optimally mapped circuits under layout restrictions, we observe a reduction up to 17\% CNOT count and 19\% CNOT depth.
\end{abstract}

\end{frontmatter}


\section{Introduction}
\label{sec:introduction}

Quantum Computing promises speedup in solving computationally hard and classically intractable problems.
Logical formulations of such problems are compiled to enable execution on quantum processors.
The Quantum compilation pipeline broadly consists of two main stages, Circuit Synthesis and Layout Synthesis.
Circuit Synthesis mainly focuses on the decomposition of abstract circuits to a target gate set.
Layout Synthesis instead focuses on satisfying hardware restrictions.
For instance, not all physical qubits interact with each other in some current quantum processors.
Thus, quantum gates that act on 2 qubits can only be scheduled on adjacent physical qubits.
In the current Noisy Intermediate Scale Quantum (NISQ) era, noise is inherent to quantum computers.
Every execution of a gate can increase the error in the computation.
For practical quantum computing, error reduction is of utmost importance.
Optimization techniques are applied throughout the compilation pipeline.
In particular, reducing gate count and circuit depth can directly impact the error rate.

While an optimal synthesis for the whole compilation pipeline is ideal, it is an extremely hard problem.
For instance, \cite{Nagarajan2021QuantumCircuitOptAO} proposes SMT-based synthesis under hardware connectivity restrictions and target gate set.
From \cite{Nagarajan2021QuantumCircuitOptAO}, it is clear that synthesis beyond 4 qubits is impractical.
Essentially, to optimize a $n$ qubit circuit one needs to consider its $2^n \times 2^n$ unitary matrix.
The alternative is to optimize error-prone gates like 1-qubit T-gates and 2-qubit CNOT-gates for error reduction.
Several approaches are applied for T-gate optimization~\cite{DBLP:journals/tcad/AmyMM14,DBLP:journals/tcad/AmyMMR13} in tools like
T-par\footnote{https://github.com/meamy/t-par} and Feynman.\footnote{https://github.com/meamy/feynman}
While such tools reduce T-gate count and depth, they can significantly increase CNOT-gate count.
As a result, CNOT optimization without changing the T-gate count has been proposed,
based on Gaussian elimination~\cite{beth2001quantum}, Greedy algorithms~\cite{10.1145/3474226,de2021reducing},
Steiner tree~\cite{DBLP:journals/qic/KissingerG20,9914638},  SAT~\cite{DBLP:conf/rc/MeuliSM18}, and ASP~\cite{DBLP:conf/cilc/PiazzaRW23,DBLP:conf/aiqxqia/PiazzaR23}.
While heuristic techniques are well studied, exact approaches are still unexplored in many variations.

\paragraph*{Contributions}
We consider two variations of CNOT synthesis, one with qubit permutation and one with CNOT restrictions.
For qubit permutation, we define weak equivalence (W) where the order of output qubits is free. This allows for more -- often smaller -- solutions than exact synthesis with strong equivalence (S).
For CNOT restrictions (R), we only allow CNOT gates on adjacent qubits, i.e., layout-aware synthesis.
We are in particular interested in 4 variants: S, S+R, W, and W+R.
Adding restrictions makes the problem NP-hard~\cite{DBLP:conf/soda/JiangSTW0Z20}, while the complexity without them is still open~\cite{DBLP:conf/soda/JiangSTW0Z20}.
We encode such hard problems into Classical Planning, Propositional satisfiability (SAT), and Quantified Boolean Formulas (QBF).
For the first time, we provide optimal encodings for W and W+R synthesis variants.
For the S, S+R and W variants, we experiment with peephole optimization on
arbitrary quantum circuits, in which individual CNOT-slices are optimized.
We validate this on standard T-gate optimal benchmarks.
We extended our open source tool Q-Synth~v3\footnote{Q-Synth v3, available at~\url{https://github.com/irfansha/Q-Synth}} to include all encoding variants of CNOT synthesis mentioned above.\footnote{Paper accepted for ECAI 2024, Santiago de Compostela, Oct 2024.}

\section{Preliminaries}
\label{sec:preliminaries}

\subsection{CNOT circuits}
\label{subsec:cnotcircuits}

In this paper, we focus on special circuits called CNOT circuits,
which consist solely of 2-qubit CNOT gates (controlled-NOT).
A CNOT gate takes two inputs, a control qubit with $a$ and a target qubit with $b$, and outputs $a \oplus b$ on the target qubit.
For example, Table~\ref{table:orcircuit} shows a CNOT circuit with 6 CNOT gates.
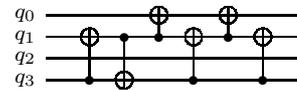
\begin{table}[htbp]
  \caption{Original CNOT Circuit}
  \label{table:orcircuit}
  \[
    \begin{array}{c}
      \Qcircuit @C=0.8em @R=0.2em @!R { \\
      \lstick{{q}_{0}} & \qw & \qw           & \qw           & \targ    & \qw           & \targ     & \qw           & \qw & \qw\\
      \lstick{{q}_{1}} & \qw & \targ         & \ctrl{2}      & \ctrl{-1}& \targ         & \ctrl{-1} & \targ         & \qw & \qw\\
      \lstick{{q}_{2}} & \qw & \qw           & \qw           & \qw      & \qw           & \qw       & \qw           & \qw & \qw\\
      \lstick{{q}_{3}} & \qw & \ctrl{-2}     & \targ         & \qw      & \ctrl{-2}     & \qw       & \ctrl{-2}     & \qw & \qw
       }
  \end{array}
  \]
\end{table}

\begin{table*}[htbp]
  \caption{Column additions for respective CNOT gates in the example circuit}
  \centering
  \label{table:matrixadditions}  
    \[
    \begin{blockarray}{cccc}
    q_0 & q_1 & q_2 & q_3 \\
    \begin{block}{(cccc)}
    1 & 0 & 0 & 0\\
    0 & 1 & 0 & 0\\
    0 & 0 & 1 & 0\\
    0 & 0 & 0 & 1\\
    \end{block}
    \end{blockarray}  
    \xrightarrow{q_3, q_1}
    \begin{pmatrix}
    1 & 0 & 0 & 0\\
    0 & 1 & 0 & 0\\
    0 & 0 & 1 & 0\\
    0 & \mathbf{1} & 0 & 1\\
    \end{pmatrix}
    \xrightarrow{q_1, q_3}
    \begin{pmatrix}
    1 & 0 & 0 & 0\\
    0 & 1 & 0 & \mathbf{1}\\
    0 & 0 & 1 & 0\\
    0 & 1 & 0 & \mathbf{0}\\
    \end{pmatrix}
    \xrightarrow{q_1, q_0}
    \begin{pmatrix}
    \cdots
    \end{pmatrix}
    \xrightarrow{q_3, q_1}
    \begin{pmatrix}
    \cdots
    \end{pmatrix}
    \xrightarrow{q_1, q_0}
    \begin{pmatrix}
    \cdots
    \end{pmatrix}
    \xrightarrow{q_3, q_1}
    \begin{blockarray}{cccc}
      q_0 & q_1 & q_2 & q_3 \\
      \begin{block}{(cccc)}
        1 & 0 & 0 & 0\\
        1 & \mathbf{1} & 0 & 1\\
        0 & 0 & 1 & 0\\
        0 & 1 & 0 & 0\\
      \end{block}
      \end{blockarray}
    \]
    \end{table*}

CNOT sub-circuits appear frequently in quantum circuits, since CNOT is the only
binary gate in many quantum platforms.
Optimizing such sub-circuits directly impacts the overall error.
Every $n$-qubit CNOT circuit can be represented by a so-called \emph{parity} matrix i.e., a full-rank $n\times n$ matrix in $\mathit{GL}_n(F_2)$~\cite{beth2001quantum,Nagarajan2021QuantumCircuitOptAO}.
Adding column $i$ to column $j$ (modulo 2) in this matrix corresponds
to applying a CNOT gate with control qubit $i$ to target qubit $j$.
So the minimal CNOT circuit corresponds to finding the shortest series of column additions
to obtain the goal matrix from the Identity matrix.
The parity matrix formulation for CNOT circuits is much more compact than the usual $2^n\times 2^n$ complex unitary matrix for arbitrary quantum circuits.
\begin{definition}
Given a CNOT circuit C on $n$ qubits, we define $\fullrankmatrix_C$ to be its parity matrix in $\mathit{GL}_n(F_2)$ generated by applying all CNOT gates in $C$ to the $n\times n$ Identity matrix.
\end{definition}

The columns of the parity matrix are labeled with the qubits.
Given a circuit, one can transform the Identity matrix to the final parity matrix by applying column additions corresponding to the CNOT gates.
For example, Table~\ref{table:matrixadditions} shows such a transformation via column additions of our example circuit.
The right-most matrix in Table~\ref{table:matrixadditions} shows the equivalent matrix for the CNOT circuit.
In the parity matrix, each column represents the output of the corresponding qubit.
For instance, column $q_2$ has the bit sequence $0, 0, 1, 0$ representing the untouched qubit $q_2$.
On the other hand, column $q_0$ with bit sequence $1, 1, 0, 0$ represents $q_0 \oplus q_1$.

\subsection{Classical planning}
\label{subsec:classicalplanning}

Given a description of a world, finding a sequence of actions that transform an initial state to some goal state is Automated Planning.
In Classical Planning~\cite{ghallab2004automated}, the actions are deterministic and there exists a single initial state.
Any reachability encoding can be elegantly encoded in such a specification.
The problem is specified using Domain and Problem files in the Planning Domain Definition Language (PDDL)~\cite{DBLP:journals/jair/FoxL03}.
A domain file defines the predicates that describe the world and lists schematic actions that can change the world.
A problem file specifies the objects used, the initial state, and the goal state.
One can then use existing State-of-the-art domain-independent planners to solve problems.
Layout Synthesis of Quantum circuits has been successfully encoded before in classical planning \cite{ShaikvdP2023}.

\subsection{Propositional satisfiability}
\label{subsec:sat}

Given a boolean formula, finding an assignment that makes the formula true is a propositional satisfiability (SAT) problem.
In recent years, many (NP-complete and NP-hard) problems have been successfully encoded and solved using SAT~\cite{DBLP:series/faia/336,DBLP:journals/cacm/FichteBHS23}.
Several synthesis-related problems in Quantum Computing have been encoded in SAT~\cite{Schneider2022ASE,Peham2023DepthOptimalSO,DBLP:conf/rc/MeuliSM18,shaik2024optimal}.
Since we are interested in optimal solutions, SAT-based solving is a promising technique for proving optimality.

\subsection{Quantified boolean formulas}
\label{subsec:qbf}

Quantified Boolean Formula (QBF) Logic~\cite{DBLP:series/faia/BeyersdorffJLS21} is an extension of propositional logic with universal and existential quantifiers.
One can encode a propositional formula in a more compact way taking advantage of inherent structure.
When propositional formulas get too large to encode, encoding in QBF is an alternative.
For instance, using QBF-based encodings helped in avoiding large Organic Synthesis encodings based on Planning in~\cite{DBLP:conf/aips/ShaikP22}.

\section{Optimal CNOT synthesis}
\label{sec:cnotsynth}

In this section, we discuss CNOT optimization and its variations with synthesis.
We will first establish different notions of equivalence between CNOT circuits.
Then we discuss layout-aware synthesis, in the presence of connectivity restrictions.
Finally, we discuss the relevant combinations of synthesis variants for the encodings in this paper.
In section~\ref{sec:relatedwork}, we present related work and compare it with our approach.

\subsection{CNOT circuit equivalence}
\label{subsec:cnotequv}

For optimal synthesis, we first need to establish equivalence between CNOT circuits.
In general, two quantum circuits are equivalent if their unitary matrices are the same.
Intuitively, equivalent circuits have the same input-output behavior.
Note that equivalent circuits can have different gate counts and circuit depths.
The general idea for optimization is to compute an equivalent circuit with either a lower gate count or circuit depth.

\paragraph*{Strong equivalence (S)}
Every parity matrix has a corresponding unitary matrix.
For CNOT circuits, we can directly use the parity matrix representation for this equivalence relation.
If two CNOT circuits have the same parity matrix, then they are strongly equivalent~\cite{beth2001quantum,Nagarajan2021QuantumCircuitOptAO}.
\begin{definition}
Two CNOT circuits $C, C'$ are \emph{strongly equivalent} if and only if $\fullrankmatrix_C = \fullrankmatrix_{C'}$.
\end{definition}
For example, consider the CNOT circuit in Table~\ref{table:orcircuit} with the final matrix in Table~\ref{table:matrixadditions}.
Synthesizing optimal column operations to reach $M_C$ is equivalent to synthesizing an optimal circuit $C'$.
One can synthesize the same parity matrix by using only three column additions on $(q_1,q_0), (q_3, q_1)$, and $(q_1, q_3)$.
Thus, the resulting equivalent circuit as in Table~\ref{table:optcircuitstrongeq} only has 3 CNOT gates (optimal) as instead of 6.

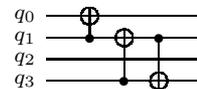
\begin{table}[htbp]
\caption{Optimized circuit via S, with 3 CNOT gates only}
\label{table:optcircuitstrongeq}
\[
\begin{array}{c}
  \Qcircuit @C=0.8em @R=0.2em @!R { \\
   \lstick{{q}_{0}} & \qw & \targ         & \qw           & \qw       & \qw & \qw \\
   \lstick{{q}_{1}} & \qw &  \ctrl{-1}    & \targ         & \ctrl{2}  & \qw & \qw \\
   \lstick{{q}_{2}} & \qw & \qw           & \qw           & \qw       & \qw & \qw \\
   \lstick{{q}_{3}} & \qw & \qw           & \ctrl{-2}     & \targ     & \qw & \qw
   }
\end{array}
\]
\end{table}

\paragraph*{Weak equivalence (W)}
While one can use strong equivalence for optimal synthesis, the definition is somewhat restrictive.
The order of input qubits and output qubits is the same in the optimized circuit in Table~\ref{table:optcircuitstrongeq}.
However, one could allow permutations of the output qubits within a circuit,
as long as one keeps track of the final ``physical'' position of the ``logical'' output qubits.
For example, Table~\ref{table:optcircuitweakeqmeas} shows an equivalent circuit where the order of output qubits is changed.
\begin{table}[htbp]
  \caption{Optimized circuit using W (weak equivalence), with 2 CNOTs. The final position of permuted output qubits are represented using circles.}
  \label{table:optcircuitweakeqmeas}
  \[
  \begin{array}{c}
    \Qcircuit @C=0.8em @R=0.2em @!R {
     \lstick{{q}_{0}} & \qw & \targ         & \qw       & \qw & \measure{q_0}\\
     \lstick{{q}_{1}} & \qw &  \ctrl{-1}    & \ctrl{2}  & \qw & \measure{q_3}\\
     \lstick{{q}_{2}} & \qw & \qw           & \qw       & \qw & \measure{q_2}\\
     \lstick{{q}_{3}} & \qw & \qw           & \targ     & \qw & \measure{q_1}
     }
  \end{array}
  \]
  \end{table}

In CNOT circuits, the output qubit permutation simply corresponds to the permutation of columns in the parity matrix.
\begin{definition}
Given a permutation $P$ and a matrix $\fullrankmatrix_C$, we define $P(\fullrankmatrix_C)$ be the column permuted matrix.
\end{definition}

Intuitively, it might be possible to reach some permutation of the matrix in fewer steps.
We now define the weak equivalence based on the permutation of matrices.

\begin{definition}
Two CNOT circuits $C$ and $C'$ are \emph{weakly equivalent} if and only if there exists a permutation $P$ such that $P(\fullrankmatrix_C) = \fullrankmatrix_{C'}$.
\end{definition}

In our example, by using weak equivalence only 2 column operations are required to reach a permuted matrix.
Table~\ref{table:optcircuitweakeqmeas} shows such an optimal circuit with only 2 CNOT gates.
We use circles to denote the permuted positions of output qubits.

To convert a weakly equivalent circuit to a strongly equivalent circuit, one can add tailing swaps to model the permutation of output qubits.
When there are no connectivity restrictions on CNOT gates, these swap gates have \emph{zero-cost}.
One can remove swaps by simply relabelling gates in constant time.
In our example, using swaps results in a strongly equivalent optimized circuit as shown in Table~\ref{table:optcircuitweakeqswaps}.

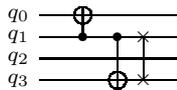
  \begin{table}[htbp]
  \caption{Optimized circuit using W with 2 CNOTs and one SWAP}
  \label{table:optcircuitweakeqswaps}
  \[
   \begin{array}{c}
    \Qcircuit @C=0.8em @R=0.2em @!R {
       \lstick{{q}_{0}} & \qw & \targ         & \qw       & \qw            & \qw & \qw \\
       \lstick{{q}_{1}} & \qw &  \ctrl{-1}    & \ctrl{2}  & \qswap \qwx[2] & \qw & \qw \\
       \lstick{{q}_{2}} & \qw & \qw           & \qw       & \qw            & \qw & \qw \\
       \lstick{{q}_{3}} & \qw & \qw           & \targ     & \qswap         & \qw & \qw
       }
  \end{array}
  \]
  \end{table}

\subsection{Restricted CNOT connections (R)}
\label{subsec:restrictedcnotconnections}

We also explore layout-aware CNOT optimization.
In some quantum platforms, not all qubit pairs are connected thus restricting 2-qubit gate execution.
One can only apply CNOT gates on adjacent pairs of qubits based on some coupling graph.
Usually, Circuit Synthesis and Layout Synthesis are separated, leading to suboptimal results.
For instance, most Layout Synthesis techniques do not take CNOT gate cancellation opportunities into account when transpiling.

We integrate these phases, by adapting our CNOT synthesis to respect connectivity constraints.
Given a restricted set of CNOT connections, we only allow column additions that correspond to adjacent qubits when synthesizing the final matrix (or a permutation of it).
We now obtain the minimal CNOT circuit that satisfies the restrictions.

Suppose, we want to optimize our example circuit in Table~\ref{table:orcircuit},
allowing CNOT gates only on qubit pairs $(0,1), (1,2)$, and $(2,3)$.
If we insist on strong equivalence, we need 8 CNOT gates (optimal) as shown in Table~\ref{table:optcircuitstrongeqcnotr}.
Note that we need more than the original 6 CNOT gates, due to connectivity restrictions.
Allowing weak equivalence requires only 5 CNOT gates (optimal) as shown in Table~\ref{table:optcircuitweakeqcnotr}.

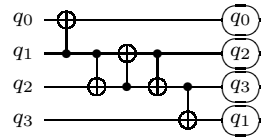
\begin{table}[htbp]
  \caption{Optimized circuit with S+R, using 8 CNOTs}
  \label{table:optcircuitstrongeqcnotr}
  \[
  \begin{array}{c}
    \Qcircuit @C=0.8em @R=0.2em @!R {
     \lstick{{q}_{0}} & \targ     & \qw       & \qw        & \qw     & \qw      & \qw     & \qw     & \qw \\
     \lstick{{q}_{1}} & \ctrl{-1} & \qw       & \targ      & \ctrl{1}& \qw      & \qw     & \ctrl{1}& \qw \\
     \lstick{{q}_{2}} & \ctrl{1}  & \targ     & \ctrl{-1}  & \targ   & \targ    & \ctrl{1}& \targ   & \qw \\
     \lstick{{q}_{3}} & \targ     & \ctrl{-1} & \qw        & \qw     & \ctrl{-1}& \targ   & \qw     & \qw
     }
  \end{array}
  \]
  \end{table}
  
  \begin{table}[htbp]
    \caption{Optimized circuit with W+R, using 5 CNOTs. The final placement of permuted output qubits are represented using circles.}
    \label{table:optcircuitweakeqcnotr}
    \[
      \begin{array}{c}
        \Qcircuit @C=0.6em @R=0.2em @!R {
         \lstick{{q}_{0}} & \targ     & \qw     & \qw        & \qw     & \qw     & \qw & \measure{q_0} \\
         \lstick{{q}_{1}} & \ctrl{-1} & \ctrl{1}& \targ      & \ctrl{1}& \qw     & \qw & \measure{q_2} \\
         \lstick{{q}_{2}} & \qw       & \targ   & \ctrl{-1}  & \targ   & \ctrl{1}& \qw & \measure{q_3} \\
         \lstick{{q}_{3}} & \qw       & \qw     & \qw        & \qw     & \targ   & \qw & \measure{q_1}
         }
      \end{array}
      \]
    \end{table}

\subsection{Metrics and relevant synthesis variants}
\label{subsec:complexityandvariants}

We consider three metrics on quantum circuits in this paper, CNOT-gate count, depth, and CNOT depth.
The {\em CNOT count} is simply the number of CNOT gates in a circuit.
The {\em depth} of a circuit is the length of the longest path in the dependency graph
of its gates, connected by direct input-output dependencies.
The {\em CNOT depth} of a circuit is the largest number of CNOT gates on any dependency chain.
For instance, the circuit 
in Table~\ref{table:optcircuitstrongeqcnotr} has CNOT count 8, but its CNOT depth
is only 7 (since the first two CNOT gates are applied in parallel).
For CNOT circuits, depth and CNOT depth always coincide. 

For CNOT and depth optimization, optimal synthesis is a computationally hard problem.
Variants with CNOT restrictions have been proven to be NP-hard for both gate count
and depth optimization~\cite{DBLP:conf/soda/JiangSTW0Z20} metrics.
In fact, for synthesis with CNOT restrictions, even finding approximate solutions is NP-Hard~\cite{iwama2002transformation}.
We encode such hard problems in Classical Planning, SAT, and QBF.
In particular, we are interested in $4$ synthesis variants:
\begin{itemize}
  \item S: Synthesis with strong equivalence (and no restrictions).
  \item W: Synthesis with weak equivalence (and no restrictions).
  \item S+R: Synthesis with strong equivalence and CNOT restrictions.
  \item W+R: Synthesis with weak equivalence and CNOT restrictions.
\end{itemize}
Not all variants can be encoded efficiently in every solving technique.
For instance, we found an efficient encoding of the W variant in SAT, 
but it seems more difficult in classical planning and QBF.
So we encode selected variants for each technique:
\begin{itemize}
  \item Classical Planning: only S and S+R with CNOT gate optimization.
  \item SAT: All $4$ variations for both CNOT count and CNOT depth.
  \item QBF: S and S+R for both CNOT count and depth optimization.
\end{itemize}  

\section{CNOT synthesis as planning}
\label{sec:cnotsynthesisplanning}

In this section, we first describe the encoding for the S+R variant in Classical Planning in PDDL.
We encode the synthesis as a reachability problem, where nodes of a graph represent the state of the matrix and edges represent column additions.
Given a circuit, we first compute its parity matrix which corresponds to the goal node.
The shortest path from the initial node with the Identity matrix to the goal node corresponds to the optimal number of CNOT gates.

All our objects, which label rows and columns, are of type \texttt{qubit}.
We use the following two predicates to represent the state:
\begin{itemize}
\item \texttt{(m ?r ?c - qubit)}: represents a matrix element with ?r row and ?c column parameters.
\item \texttt{(connected ?a ?b - qubit)}: static predicate to represent connected qubit parameters ?a and ?b.
\end{itemize}
To apply a CNOT gate on two qubits, we encode the corresponding column addition as an action.
In preconditions, we specify that a CNOT gate can be applied only on different qubits and the qubits must be connected.
In effects, for each row, the element in the target column is flipped if the control column element is true.
We use conditional effects in PDDL to encode the effects.
Listing~\ref{lst:domain} is the corresponding domain file with all predicates and actions in PDDL.
\begin{lstlisting}[caption={Domain for S+R CNOT synthesis in PDDL Format},label={lst:domain},
  language=stripspddl,basicstyle=\small,mathescape]
(:predicates
  (m ?r ?c - qubit)(connected ?a ?b - qubit))
(:action cnot
  :parameters (?c ?t - qubit)
  :precondition (and 
    (not(= ?c ?t))(connected ?c ?t))
  :effect (and 
    (forall(?r - qubit)
      (when (and (m ?r ?c)    (m ?r ?t))
                          (not(m ?r ?t))))
    (forall(?r - qubit)
      (when (and (m ?r ?c)(not(m ?r ?t)))
                              (m ?r ?t)))))
\end{lstlisting}
For any CNOT synthesis instance, one can use the same Domain file.
The Problem file, on the other hand, defines instance-specific information i.e., objects, initial and goal states.
Listing~\ref{lst:problemsnippet} shows snippets of the problem file for our example (see Table~\ref{table:orcircuit}).
For CNOT synthesis, we define one object per qubit.
In the initial state, we encode the Identity matrix i.e., only diagonal elements are set to true.
We specify which qubits are connected based on an input coupling graph.
In the initial state specification, one only specifies the true propositions, and unspecified propositions are negated by default.
For the goal state, we encode the final matrix for the given circuit.
\begin{lstlisting}[caption={Problem snippets in PDDL for the example circuit},label={lst:problemsnippet},
  language=stripspddl,basicstyle=\small,mathescape]
(:objects $q_0$ $q_1$ $q_2$ $q_3$ - qubit)
(:init
  (m $q_0$ $q_0$)(m $q_1$ $q_1$)(m $q_2$ $q_2$)(m $q_3$ $q_3$)
  (connected $q_0$ $q_1$)(connected $q_1$ $q_0$)
  (connected $q_1$ $q_2$)(connected $q_2$ $q_1$)
  (connected $q_2$ $q_3$)(connected $q_3$ $q_2$))
(:goal (and
      (m $q_0$ $q_0$) $\dots$(not(m $q_0$ $q_2$))(not(m $q_0$ $q_3$))
      (m $q_1$ $q_0$) $\dots$(not(m $q_1$ $q_2$))            (m $q_1$ $q_3$)
  (not(m $q_2$ $q_0$))$\dots$         (m $q_2$ $q_2$) (not(m $q_2$ $q_3$))
  (not(m $q_3$ $q_0$))$\dots$(not(m $q_3$ $q_2$))(not(m $q_3$ $q_3$))))
\end{lstlisting}
An optimal plan i.e., a plan with minimal actions, corresponds to an optimal circuit.
We can then use any off-the-shelf optimal planners to synthesize optimal CNOT circuits.
One could also use heuristic planners for fast synthesis in case of large instances.
For S synthesis without restrictions, we simply drop the \texttt{connected} predicate from the domain and problem files.

\section{CNOT synthesis as SAT}
\label{sec:satencoding}

Encoding in classical planning is elegant and easy to understand.
However, classical planners are good at finding fast heuristic plans but face scalability issues for computing optimal plans.
Since our synthesis problem is encoded as a bounded reachability problem, a SAT encoding for optimal synthesis is promising.

\subsection{Gate optimal encoding}
\label{subsec:cnotcountoptimal}
For CNOT gate optimality, we apply a standard one-hot reachability encoding.
First, we define variables for matrices which represent the state at each time step.
We represent the matrix element in row $r$ and column $c$ at time step $t$ as $\matrixvar_{r,c}^{t}$.
At each time step, we apply a single column addition on some control and target columns.
We represent the control column as $\ctrlvar^t$ and the target column as $\trgtvar^t$ at time step $t$.
For a plan length of $k$, we define $k$ copies of action variables and $k+1$ copies of state variables.
\paragraph*{S+R synthesis}
The initial state corresponds to the Identity matrix.
We encode it using Exactly-One (EO) constraints on row elements and unit clauses for diagonal elements.
\begin{equation}
\bigwedge_{r=0}^{\numqubits-1} \EO(\matrixvar_{r,0}^{0}, \cdots, \matrixvar_{r,\numqubits-1}^{0}) \land \bigwedge_{q=0}^{\numqubits-1} \matrixvar_{q,q}^{0}
\end{equation}
For each transition step, exactly one control and target column is chosen (see Equation 2).
Given a coupling graph with a set of connected qubit pairs $\cppairs$, we only allow corresponding column pairs (see Equation 3).
For state updates, we encode constraints:
\begin{itemize}
\item For every row, we update target column matrix variables based on control column matrix variables. If the control variable is:
\begin{itemize}
  \item true, then the target variable is flipped (see Equation 4)
  \item false, then the target variable is propagated (see Equation 5)
\end{itemize}
\item All untouched column variables are propagated (see Equation 6).
\end{itemize}
For time steps $t \in \{0,\cdots,k-1\}$, we specify:
\begin{align}
&\EO(\ctrlvar^t_{0}, \cdots, \ctrlvar^{t}_{\numqubits-1}) \land \EO(\trgtvar^t_{0},\cdots, \trgtvar^{t}_{\numqubits-1}) \\
&\bigwedge_{i=0}^{\numqubits-1}\bigwedge_{j=0}^{\numqubits-1} (\{\neg \ctrlvar^{t}_{i} \lor \neg \trgtvar^{t}_{j} \mid (i,j) \notin \cppairs\})\\
&\bigwedge_{(i,j)\in \cppairs} \bigwedge_{r=0}^{\numqubits-1} (\ctrlvar^{t}_{i} \land \trgtvar^{t}_{j} \land \matrixvar_{r,i}^{t}) \implies (\matrixvar_{r,j}^{t} \neq \matrixvar_{r,j}^{t+1})\\
&\bigwedge_{(i,j)\in \cppairs} \bigwedge_{r=0}^{\numqubits-1} (\ctrlvar^{t}_{i} \land \trgtvar^{t}_{j} \land \neg \matrixvar_{r,i}^{t}) \implies (\matrixvar_{r,j}^{t} = \matrixvar_{r,j}^{t+1})\\
&\bigwedge_{i=0}^{\numqubits-1} \bigwedge_{r=0}^{\numqubits-1} \neg \trgtvar^{t}_i \implies (\matrixvar_{r,i}^{t} = \matrixvar_{r,i}^{t+1})
\end{align}
For a given circuit $C$, we encode the goal state with the corresponding final matrix $\fullrankmatrix_C$.
For every $1$ in the matrix, we add positive unit clauses in the goal state matrix and negative ones for every $0$ (see Equation 7).
In synthesis variant S, all different qubit pairs are connected.
\begin{equation}
\bigwedge_{i=0}^{\numqubits-1}\bigwedge_{j=0}^{\numqubits-1} \Big(\bigwedge_{\fullrankmatrix_C[i,j] = 1} \matrixvar^{k}_{i,j} \land \bigwedge_{\fullrankmatrix_C[i,j] = 0} \neg \matrixvar^{k}_{i,j}\Big)
\end{equation}

\paragraph*{W synthesis}
One can encode column permutation of goal matrix for weak equivalence.
However, such a permutation would result in many clauses.
Instead, we observe that every circuit with permuted output qubits has an equivalent circuit with permuted input qubits.
An input-permuted circuit has swaps at the start instead of the end of the circuit.
Removing initial swaps simply relabels CNOT gates and results in the permutation of output qubits.
Note that such relabelling does not change the number of CNOTs in the circuit.
For example, Table~\ref{table:optcircuitweakeqinitswaps} shows an example circuit with initial swaps instead of the end as in Table~\ref{table:optcircuitweakeqswaps}.
\begin{table}[htbp]
  \caption{Optimized circuit with initial swaps via W, with 2 CNOTs}
  \label{table:optcircuitweakeqinitswaps}
  \[
   \begin{array}{c}
    \Qcircuit @C=0.8em @R=0.2em @!R { \\
       \lstick{{q}_{0}} & \qw            & \targ         & \qw       & \qw & \qw \\
       \lstick{{q}_{1}} & \qswap \qwx[2] & \qw           & \targ     & \qw & \qw \\
       \lstick{{q}_{2}} & \qw            & \qw           & \qw       & \qw & \qw \\
       \lstick{{q}_{3}} & \qswap         & \ctrl{-3}     & \ctrl{-2} & \qw & \qw
       }
  \end{array}
  \]
  \end{table}
Identity matrix permutation can be encoded elegantly using exactly-one constraints on the time step $0$ state variables.
This can be achieved by dropping unit clauses in Equation 1 and adding exactly-one constraints on column variables.
Essentially, we replace Equation 1 with Equation 8.
\begin{equation}
\bigwedge_{r=0}^{\numqubits-1} \EO(\matrixvar_{r,0}^{0}, \cdots, \matrixvar_{r,\numqubits-1}^{0}) \land \bigwedge_{c=0}^{\numqubits-1} \EO(\matrixvar_{0,c}^{0}, \cdots, \matrixvar_{\numqubits-1,c}^{0})
\end{equation}

\paragraph*{W+R synthesis}
In the presence of CNOT restrictions, removing initial swaps can result in CNOT gates applied on restricted qubit pairs.
To circumvent this problem, we encode symbolic qubit pair restrictions based on the initial permutation.
Exactly-one constraints in the Initial matrix encodes the permutation.
If $m_{i,p}^{0}$ is true then it implies that qubit $i$ is mapped to qubit $p$.
We use such information to specify the restricted qubit pairs after the permutation.
Essentially, if a restricted qubit pair $(i,j)$ is mapped to $(p,q)$ then $(p,q)$ is restricted.
We replace Equation~3 with Equation 9.
\begin{align}
\bigwedge_{i=0}^{\numqubits-1}\bigwedge_{j=0}^{\numqubits-1}\bigwedge_{p=0}^{\numqubits-1}\bigwedge_{q=0}^{\numqubits-1}  &(\{m_{i,p}^{0} \land m_{j,q}^{0} \nonumber \\
&\implies  \neg \ctrlvar^{t}_{p} \lor \neg \trgtvar^{t}_{q} \mid (i,j) \notin \cppairs\})
\end{align}

\subsection{Depth optimal encoding with parallel plans}
\label{subsec:cnotdepthoptimal}

CNOT depth is another important metric in the optimization of quantum circuits.
CNOTs acting on different qubits can be applied at the same depth.
Depth-based synthesis can be encoded in SAT by allowing parallel CNOTs at each time step.
The makespan of such an encoding corresponds to the depth of the synthesized circuit.
We only discuss the S+R synthesis variant; the other 3 variants directly follow from the above gate-optimal encoding.
Similar to the gate-optimal encoding, we define the same matrix variables to represent the state.
Both the initial and goal constraints are exactly the same, i.e., Equations 1 and 7 stay the same.
To allow parallel CNOT gates, we define one variable $\cnotvar_{i,j}^{t}$ for each qubit pair $(i,j)$ at time step $t$.
To respect CNOT restrictions, we disable the CNOT variables on restricted pairs (see Equation~10).
We also use target variables $\trgtvar^t_q$ as before for propagation of untouched column variables.
So propagation constraints as in Equation 6 stay the same for depth optimal encoding, now Equation 16.
Note that multiple target columns can be changed due to parallel CNOTs.
We handle parallel CNOT operations by specifying (see Equations~11-15):
\begin{itemize}
\item Atmost-One (AMO) CNOT gate is applied on a qubit.
\item Atleast-One (ALO) CNOT gate is applied at each time step. Only for efficiency, dropping them would not affect correctness.
\item Target column $\trgtvar_j$  is set to true iff some CNOT on $(i,j)$ is true.
\item For every CNOT variable and every row, we update target column variables based on control column variables:
\begin{itemize}
  \item if the control variable is true, then the target variable is flipped
  \item if the control variable is false, the target variable is propagated
\end{itemize}
\end{itemize}
For time steps $t \in \{0,\cdots,k-1\}$, we specify:
\newline
\newline
\begin{align}
& \bigwedge_{i=0}^{\numqubits-1}\bigwedge_{j=0}^{\numqubits-1} (\{\neg\cnotvar_{i,j}|(i,j) \notin \cppairs\})\\
& \bigwedge_{q=0}^{\numqubits-1} \AMO(\{\cnotvar_{i,j}^{t} \mid (i=q \text{ or } j=q \text{ and } (i,j) \in \cppairs) \})\\
& \ALO(\{\cnotvar_{i,j}^{t} \mid (i,j) \in \cppairs\})\\ 
& \bigwedge_{i}^{\numqubits-1}\big((\bigvee_{j=0}^{\numqubits-1} \cnotvar^{t}_{i,j}) = \trgtvar^{t}_{j} \big)\\
&\bigwedge_{(i,j)\in \cppairs} \bigwedge_{r=0}^{\numqubits-1} (\cnotvar^{t}_{i,j} \land \matrixvar_{r,i}^{t}) \implies (\matrixvar_{r,j}^{t} \neq \matrixvar_{r,j}^{t+1})\\
&\bigwedge_{(i,j)\in \cppairs} \bigwedge_{r=0}^{\numqubits-1} (\cnotvar^{t}_{i,j} \land \neg \matrixvar_{r,i}^{t}) \implies (\matrixvar_{r,j}^{t} = \matrixvar_{r,j}^{t+1})\\
&\bigwedge_{i=0}^{\numqubits-1} \bigwedge_{r=0}^{\numqubits-1} \neg \trgtvar^{t}_i \implies (\matrixvar_{r,i}^{t} = \matrixvar_{r,i}^{t+1})
\end{align}

\section{CNOT synthesis as QBF}
\label{sec:qbfencoding}

Even for the simplest synthesis variant S, the SAT encoding uses $O(n^2)$ variables and $O(n^3)$ clauses.
For moderately large $n$ the encoding sizes can get massive.
In CNOT synthesis, the column updates are the same for every row.
One can use universal quantification in QBF to capture this structure and generate a compact encoding.
While QBF solvers are not as mature as SAT solvers, in some cases well-structured QBF encodings can help.
In this section, we focus on the S+R synthesis variant for CNOT count optimization.
The other variants (S for CNOT count and S, S+R for CNOT depth) follow directly.
We drop W and W+R variants for QBF, as encoding column permutation symbolically is difficult.

The action variables, i.e., control and target variables are the same as in the SAT encoding.
Instead of defining column matrix variables for each row, we define a symbolic row with universal variables.
We use binary encoding for the universal variables, we define $\forallrow$ as $\{\forallrow_0, \cdots, \forallrow_{\ceil{\log(\numqubits)}-1}\}$.
For better propagation, we add one-hot encoding for symbolic row variables with existential variables $\rowvar_0, \cdots, \rowvar_{\numqubits-1}$.
The idea is to set the existential variables based on binary row variables.
We can directly use existential row variables for state update constraints similar to our SAT encoding.
We only need one set of column matrix variables to represent the complete matrix: $\colmatrixvar_{i}$ represents the $i$th column variable
(for the symbolic row $R$).

\noindent We define the prefix of our QBF encoding as follows:
\begin{align}
& \exists \ctrlvar^{0}_{0}, \cdots, \ctrlvar^{0}_{\numqubits-1} \exists \trgtvar^{0}_{0} \cdots, \trgtvar^{0}_{\numqubits-1}\\
&\cdots \\
& \exists \ctrlvar^{k-1}_{0}, \cdots, \ctrlvar^{k-1}_{\numqubits-1} \exists \trgtvar^{k-1}_{0}, \cdots, \trgtvar^{k-1}_{\numqubits-1}\\
& \forall \forallrow \quad \exists \rowvar_0, \cdots, \rowvar_{\numqubits-1}\\
& \exists \colmatrixvar^{0}_{0}, \dots , \colmatrixvar^{0}_{\numqubits-1} \cdots \exists \colmatrixvar^{k}_{0}, \dots , \colmatrixvar^{k}_{\numqubits-1}
\end{align}
First, we imply existential row variables from binary-encoded symbolic row variables.
Exactly one existential row variable is true.
\begin{equation}
\bigwedge_{i=0}^{\numqubits-1} (\bin(\forallrow,i) \implies \rowvar_{i})\quad \land \quad \EO(\rowvar_0,\cdots,\rowvar_{\numqubits-1})
\end{equation}
For the initial state, we encode the identity matrix where only diagonal matrix variables are true.
\begin{equation}
\bigwedge_{i=0}^{\numqubits-1} \rowvar_{i} = \colmatrixvar^{0}_{i}
\end{equation}  
For the goal state, we encode the final matrix $\fullrankmatrix_C$ again using existential row variables.
\begin{equation}
\bigwedge_{i=0}^{\numqubits-1}\bigwedge_{j=0}^{\numqubits-1} \bigwedge_{\fullrankmatrix_C[i,j] = 1} \rowvar_i \implies \colmatrixvar^{k}_{j} \land \bigwedge_{\fullrankmatrix_C[i,j] = 0} \rowvar_i \implies \neg \colmatrixvar^{k}_{j}
\end{equation}
Transition constraints are similar to those in our SAT encoding (see Equations 2 to 6), but here we simply drop the row indices from the SAT encoding.
For time steps $t \in \{0,\cdots,k-1\}$, we specify:
\begin{align}
&\EO(\ctrlvar^t_{0}, \cdots, \ctrlvar^{t}_{\numqubits-1}) \land \EO(\trgtvar^t_{0},\cdots, \trgtvar^{t}_{\numqubits-1})\\
&\bigwedge_{i=0}^{\numqubits-1}\bigwedge_{j=0}^{\numqubits-1} (\{\neg \ctrlvar^{t}_{i} \land \neg \trgtvar^{t}_{j} \mid (i,j) \notin \cppairs\})\\
&\bigwedge_{(i,j)\in \cppairs} (\ctrlvar^{t}_{i} \land \trgtvar^{t}_{j} \land \colmatrixvar_{i}^{t}) \implies (\colmatrixvar_{j}^{t} \neq \colmatrixvar_{j}^{t+1})\\
&\bigwedge_{(i,j)\in \cppairs} (\ctrlvar^{t}_{i} \land \trgtvar^{t}_{j} \land \neg \colmatrixvar_{i}^{t}) \implies (\colmatrixvar_{j}^{t} = \colmatrixvar_{j}^{t+1})\\
& \bigwedge_{i=0}^{\numqubits-1} \neg \trgtvar^{t}_i \implies (\colmatrixvar_{i}^{t} = \colmatrixvar_{i}^{t+1})
\end{align}

\section{Implementation and evaluation}
\label{sec:implementation}

We are mainly interested in evaluating the following aspects:
\begin{itemize}
  \item The quality improvement due to qubit permutation (W vs S).
  \item The overhead of imposing connectivity restrictions (S+R, W+R).
  \item The performance of the Planning, SAT, and QBF techniques.
\end{itemize}
\paragraph*{Peephole optimization}
To allow CNOT optimization in arbitrary circuits, we employ \emph{Peephole optimization} using a standard \emph{slice-and-replace} approach.
Given a quantum circuit in QASM format,
we extract the CNOT slices from the circuit's dependency DAG as follows:
We start from the top, such that each slice has one maximal CNOT sub-circuit followed by an arbitrary number of non-CNOT gates.
For each CNOT sub-circuit, we optimize its gate count or depth.
Finally, we replace each CNOT sub-circuit with its optimal counterpart.

Once the slicing is fixed, the order in which slices are treated does not matter for S, S+R, and W variants.
Furthermore, the optimal number of CNOTs is fixed for a given slicing.
The W+R encoding cannot be used directly in our peephole optimization.
Since a permutation in one slice can break CNOT connections in subsequent slices, the order of slice optimization matters.
While solving slice-by-slice from top to bottom gives correct results, the final CNOT count might be sub-optimal, even for the given slicing.
Hence, we do not apply peephole optimization with the W+R variant in this paper.

\subsection{Experimental setup}
\label{subsec:experimentalsetup}
Our tool Q-Synth v1\footnote{\url{https://github.com/irfansha/Q-Synth/releases/tag/Q-Synth-v1.0-ICCAD23}} and v2\footnote{https://github.com/irfansha/Q-Synth/releases/tag/Q-Synth-v2.0-SAT2024} solve layout synthesis using classical planning and SAT solving, respectively.
We extended Q-Synth to solve CNOT synthesis with Planning, SAT, and QBF.
We provide an open-source tool Q-Synth v3\footnote{\url{https://github.com/irfansha/Q-Synth/releases/tag/Q-Synth-v3.0-ECAI24}} that implements all encoding variants discussed, including peephole optimization.
For experimental evaluation, we consider standard T-gate optimized benchmarks~\cite{10.1145/3474226} generated by T-par.
We consider all benchmarks with up to 14 qubit circuits and at most 200 CNOT gates resulting in 11 instances.
We propose two experiments to address our research questions.
\paragraph*{Experiment 1}
We optimize CNOT count and depth on the benchmarks with S variant encodings of classical planning (CP), SAT, and QBF.
To investigate the impact of qubit permutation, we compare S encodings with W encodings in SAT.
Further, we compare our results with the state-of-the-art heuristic CNOT optimization tool DaCSynth (DS) in both gate~\cite{10.1145/3474226} and depth~\cite{de2021reducing} optimization.
DaCSynth applies the same slice-and-replace approach with greedy heuristic algorithms for CNOT optimization.
We use the results from Table~2 in~\cite{10.1145/3474226} and Table~3 in~\cite{de2021reducing} on our benchmarks for a fair comparison.
Since DaCSynth is not an open-source tool, we can only compare the reported CNOT count and depth but not time and memory costs.
\paragraph*{Experiment 2}
To investigate the overhead of connectivity restrictions, we take the W-optimized circuits from Experiment 1.
We optimally map the circuits with Q-Synth~v2~\cite{shaik2024optimal} onto the 14-qubit platform IBM Melbourne.
Q-Synth~v2 maps the circuits by inserting the optimal number of swaps. We then apply S+R optimization to the result.
Since the input circuits are already optimized with W, any reduction in CNOT count or CNOT depth with S+R is significant.
\paragraph*{Tools and resources}
For CP, we use the state-of-the-art optimal planner FastDownward~\cite{helmert2011fast} with merge-and-shrink (fd-ms) heuristic.
Among the optimal planners that handle conditional effects, fd-ms performed the best in our preliminary experiments.
In the case of SAT-based solving we use Cadical-1.53~\cite{BiereFazekasFleuryHeisinger-SAT-Competition-2020-solvers} as SAT solver, and CAQE~\cite{RabeT2015} with Bloqqer preprocessor~\cite{bloqqer} as QBF solver.
In both the above experiments, for each slice in the peephole optimization, we give 600 seconds time and 8 GB memory limits.
If a timeout occurs we leave the unoptimized slice untouched.
All computations for the experiments are run on a cluster.%
\footnote{\url{http://www.cscaa.dk/grendel},
Huawei FusionServer Pro V1288H V5, with 384 GB main memory, using one 3.0 GHz Intel Xeon Gold 6248R core.}

\begin{table*}[htbp]
  \caption{Experiment 1: S vs W variants peephole synthesis on T-gate optimal circuits.}
  \label{tb:experiment1a}
  \centering
  \begin{tabular}{lrrrrrrrrrrr}
    \toprule
    &  & \multicolumn{6}{c}{CNOT Optimization} & \multicolumn{4}{c}{Depth Optimization}\\
    \cmidrule(lr){3-8}\cmidrule(lr){9-12}
    & & \multicolumn{5}{c}{CNOT count} & Depth & \multicolumn{4}{c}{Depth}\\
    \cmidrule(lr){3-7}\cmidrule(lr){8-8}\cmidrule(lr){9-12}
    Circuit (\#CNOTs/Depth)& \#n & CP(S)&SAT(S)& QBF(S)& SAT(W)   & DS~\cite{10.1145/3474226} & SAT(W) & SAT(S) & QBF(S) & SAT(W)         & DS~\cite{de2021reducing}\\
    \midrule
    barencotof3 (52/60)     & 5  & 41  & 41  & 41  & 26           & 26              & 35          & 46  & 45  & 35          & 35              \\
    barencotof4 (96/96)     & 7  & 87  & 87  & 87  & \textbf{48}  & 50              & \textbf{60} & 85  & 84  & 73          & 61              \\
    barencotof5 (134/123)   & 9  & 118 & 118 & 118 & \textbf{71}  & 73              & 87          & 118 & 114 & 108          & 87              \\
    mod54       (48/57)     & 5  & 42  & 42  & 42  & 32           & 32              & 41          & 48  & 49  & 41          & \textbf{40}     \\
    modmult55   (106/75)    & 9  & 82  & 82  & 82  & \textbf{71}  & 73              & 50          & 54  & 56  & 53          & 50              \\
    qft4        (96/185)    & 5  & 84  & 84  & 84  & 57           & \textbf{56}     & \textbf{147}& 172 & 172 & 164         & 149             \\
    rcadder6    (165/157)   & 14 & 141 & 141 & 141 & \textbf{94}  & 100             & 95          & 129 & 129 & 95          & 95              \\
    tof3        (35/46)     & 5  & 30  & 30  & 30  & \textbf{19}  & 21              & \textbf{29} & 41  & 41  & 32          & 31              \\
    tof4        (63/71)     & 7  & 55  & 55  & 55  & 37           & 37              & 45          & 59  & 57  & 56          & \textbf{43}      \\
    tof5        (97/104)    & 9  & 81  & 81  & 81  & 50           & 50              & \textbf{62} & 78  & 80  & 86          & 63               \\
    vbeadder3   (120/88)    & 10 & 86  & 86  & 86  & \textbf{53}  & 61              & 48          & 66  & 68  & 49          & \textbf{45}      \\
    \cmidrule(lr){3-7} \cmidrule(lr){8-12}
    Mean Reduction(\%)     &     & 16.3& 16.3& 16.3 &\textbf{44.9}& 43.8            & \textbf{34.2}& 15.6&15.7&  22.1       & \textbf{34.2}\\
    Max Reduction(\%)      &     & 28.3& 28.3& 28.3 &\textbf{55.8}& 50.0            & 45.5         & 28.0&25.3&  44.3       & \textbf{48.9}\\
    \bottomrule
  \end{tabular}
  \end{table*}

\paragraph*{Metrics for comparison}
We report and compare techniques on three metrics, CNOT count, depth, and CNOT depth.
Two-qubit gates are more error-prone than 1-qubit gates, so 
tools like TKET~\cite{Sivarajah_2021} and T-par~\cite{10.1145/3474226} 
mainly focus on CNOT depth.
For Experiment 1, we compare with circuit depth as only circuit depths are reported in DaCSynth paper~\cite{de2021reducing}.
For Experiment 2, we report and compare the CNOT depth for our different techniques.

\subsection{Results and discussion}
\label{subsec:resultsanddiscussion}

\paragraph*{Experiment 1}
Table~\ref{tb:experiment1a} shows the data on Experiment 1.
Under ``CNOT optimization'', we report the CNOT count for Planning, SAT, and QBF.
Here all three techniques with S synthesis performed similarly.
While the CNOT reduction is the same, classical planning had 4 timeout slices whereas SAT and QBF based optimization had 3 timeout slices.
Comparing S and W, we observe that SAT with W synthesis results in significantly more reduction (up to 55.8\%).
Since the T-par tool adds additional CNOT gates to route T gates for optimization~\cite{DBLP:journals/tcad/AmyMM14}, permuting qubits can avoid such extra CNOTs.
SAT encoding with W optimally solved all slices, thus the reported results are optimal for the given circuit slicing.

In comparison with DaCSynth (column DS), SAT with W synthesis performs well and guarantees the optimal CNOT count.
We indeed report better CNOT count compared to DaCSynth.
Surprisingly, we observed one instance (\texttt{qft4}) where DaCSynth reports a lower CNOT count.
Either the slicing in DaCSynth is different or their reported count is a mistake.

Under ``Depth Optimization'', all CNOT slices are replaced by depth-optimal slices
in all our variants.
But note that, even though CNOT slices have optimal depth locally, the global circuit depth need not be optimal for a given slicing.
Surprisingly, we observed that CNOT count optimization results in overall better depth (rightmost column under ``CNOT Optimization'').
The SAT(W) variant with CNOT optimization results in a mean depth reduction of 34.2\% (only 21.1\% with Depth optimization).
We observed that local depth optimization adds extra parallel CNOTs, thus resulting in a higher global depth.
The mean depth reduction achieved by Q-Synth and DaCSynth is the same (34.2\%).

\paragraph*{Experiment 2}
Table~\ref{tb:experiment2a} reports the results of Experiment 2.
With S+R synthesis, we observe CNOT count reduction (up to 17.1\%) in 9 out of 11 already optimally mapped circuits.
Only classical planning reported 1 timeout slice.
For SAT and QBF, the CNOT reduction we report is optimal for the given slicing.

In the case of depth optimization, we observe CNOT depth reduction (up to 11.9\%) in 4 out of 11 instances.
In Experiment 1, we observed that CNOT optimization results in better global CNOT depth reduction.
Similarly, as reported in Table~\ref{tb:experiment2a}, all three techniques report better CNOT depth reduction with local CNOT optimization.

\begin{table*}[htbp]
  \caption{Experiment 2: S+R variant peephole synthesis for optimally mapped circuits on to 14-qubit Melbourne platform.}
  \label{tb:experiment2a} 
 \centering
  \begin{tabular}{lrrrrrrrrrrr}
    \toprule
                &  \multicolumn{6}{c}{CNOT Optimization} & \multicolumn{3}{c}{Depth Optimization}\\
    \cmidrule(lr){2-7} \cmidrule(lr){8-10}
                &  \multicolumn{3}{c}{CNOT count}  &  \multicolumn{3}{c}{CNOT depth} & \multicolumn{3}{c}{CNOT depth}\\
    \cmidrule(lr){2-4}\cmidrule(lr){5-7} \cmidrule(lr){8-10}
    Circuit (\#CNOTs/CNOT depth)             & CP  & SAT          & QBF          & CP & SAT & QBF && SAT         & QBF\\
    \midrule
    barencotof3 (44/41)  & 39  & 39           & 39           & 35          & 35          & 35          && 41          & 41\\
    barencotof4 (78/70)  & 74  & 74           & 74           & 66          & 66          & 66          && 70          & 70\\
    barencotof5 (110/95) & 108 & 108          & 108          & 93          & 93          & 93          && 95          & 95\\
    mod54 (56/49)        & 48  & 48           & 48           & 43          & \textbf{40} & \textbf{40} && 45          & 45\\
    modmult55 (131/79)   & 117 & \textbf{115} & \textbf{115} & \textbf{69} & 71          & 71          && 76          & 78\\
    qft4 (105/101)       & 87  & 87           & 87           & \textbf{82} & 83          & 83          && 89          & 89\\
    rcadder6 (145/102)   & 137 & 137          & 137          & 99          & 98          & \textbf{97} && 102         & 102\\
    tof3 (34/33)         & 34  & 34           & 34           & 33          & 33          & 33          && 33          & 33\\
    tof4 (61/55)         & 61  & 61           & 61           & 55          & 55          & 55          && 55          & 55\\
    tof5 (80/71)         & 77  & 77           & 77           & 67          & \textbf{66} & 67          && 71          & 71\\
    vbeadder3 (86/75)    & 79  & 79           & 79           & 67          & 67          & 67          && 74          & 74\\
    \cmidrule(lr){2-4} \cmidrule(lr){5-10}
    Mean reduction(\%) & 7.4 & \textbf{7.6} & \textbf{7.6} & 8.0           & \textbf{8.3} & \textbf{8.3} && 2.6 & 2.3\\
    Max reduction(\%)  & 17.1 & 17.1        & 17.1         & \textbf{18.8} & 18.4         & 18.4         && 11.9 & 11.9\\
    \bottomrule
  \end{tabular}
\end{table*}

\begin{table*}[t]
  \caption{Time t (in seconds) and Memory m (in MB; -- means negligible) taken with CNOT optimization.}
  \label{tb:stats1} 
  \centering
   \begin{tabular}{lrrrrrrrrrrrrrrrr}
     \toprule
     & \multicolumn{8}{c}{Experiment 1 (S vs W)}&  \multicolumn{6}{c}{Experiment 2 (S+R)}\\
     \cmidrule(lr){2-9}\cmidrule(lr){10-15}
     & \multicolumn{2}{c}{CP(S)} & \multicolumn{2}{c}{SAT(S)} & \multicolumn{2}{c}{QBF(S)} & \multicolumn{2}{c}{SAT(W)}& \multicolumn{2}{c}{CP} & \multicolumn{2}{c}{SAT} & \multicolumn{2}{c}{QBF}\\
     \cmidrule(lr){2-3}\cmidrule(lr){4-5}\cmidrule(lr){6-7}\cmidrule(lr){8-9}\cmidrule(lr){10-11}\cmidrule(lr){12-13}\cmidrule(lr){14-15}
      Circuit     & t   & m   & t & m & t & m & t & m & t   & m   & t & m & t & m\\
      \midrule
      barencotof3 & 7    & --   & 5   & --  & 7   & --  & 5  & -- & 343 & 247  & 9   & --  & 14  & --\\
      barencotof4 & 608  & 6090 & 303 & 131 & 307 & 159 & 6  & -- & 588 & 247  & 10  & --  & 19  & --\\
      barencotof5 & 41   & 299  & 8   & --  & 22  & --  & 6  & -- & 759 & 241  & 11  & --  & 23  & --\\
      mod54       & 7    & --   & 5   & --  & 7   & --  & 5  & -- & 338 & 233  & 10  & --  & 17  & --\\
      modmult55   & 11   & --   & 23  & --  & 51  & 91  & 11 & -- & 956 & 2950 & 265 & 141 & 576 & 179\\
      qft4        & 10  & --   & 10  & --  & 19  & --  & 5  & -- & 628 & 237  & 11  & --  & 24  & --\\
      rcadder6    & 1588 & 2830 & 635 & 195 & 652 & 238 & 66 & 110 & 953 & 225  & 143 & 110 & 161 & 130\\
      tof3        & 13   & --   & 11  & --  & 6   & --  & 5  & -- & 301 & 241  & 9   & --  & 13  & --\\
      tof4        & 11   & --   & 6   & --  & 13  & --  & 5  & -- & 368 & 244  & 10  & --  & 18  & --\\
      tof5        & 612  & 5840 & 609 & 187 & 611 & 204 & 6  & -- & 524 & 227  & 10  & --  & 21  & --\\
      vbeadder3   & 621  & 4390 & 607 & 186 & 619 & 217 & 7  & -- & 434 & 247  & 10  & --  & 23  & --\\
     \bottomrule
   \end{tabular}
  \end{table*}

\begin{table}[t]
\caption{Time taken in seconds for Depth optimization.}
\label{tb:stats2}
  \centering
   \begin{tabular}{lrrrrr}
     \toprule
      & \multicolumn{3}{c}{Experiment 1}&  \multicolumn{2}{c}{Experiment 2}\\
      \cmidrule(lr){2-4}\cmidrule(lr){5-6}
      Circuit     & SAT(S) & QBF(S) & SAT(W) & SAT & QBF\\
      \midrule
      barencotof3 & 5  & 6  & 5  & 9  & 10\\
      barencotof4 & 7  & 15 & 5  & 10 & 12\\
      barencotof5 & 10 & 26 & 6  & 10 & 13\\
      mod54       & 5  & 6  & 5  & 9  & 12\\
      modmult55   & 6  & 15 & 9  & 11 & 18\\
      qft4        & 5  & 10 & 5  & 10 & 16\\
      rcadder6    & 33 & 98 & 14 & 11 & 15\\
      tof3        & 5  & 6  & 5  & 9  & 10\\
      tof4        & 6  & 13 & 5  & 10 & 12\\
      tof5        & 9  & 24 & 6  & 10 & 13\\
      vbeadder3   & 10 & 32 & 6  & 10 & 14\\
     \bottomrule
   \end{tabular}
  \end{table}

\paragraph*{SAT vs QBF efficiency}
Tables~\ref{tb:stats1} and~\ref{tb:stats2} show the time and memory taken by all our encodings.
For CNOT optimization with S (Table~\ref{tb:stats1}, Experiment 1), we observe that SAT and QBF techniques perform similarly in terms of time and memory.
In most cases, Cadical (SAT) is slightly faster and takes less memory than CAQE (QBF).
Interestingly, on a large slice from the 14-qubit circuit, CAQE is slightly faster than Cadical.
This can happen because the QBF encoding is only linear in variables and quadratic in constraints, while our SAT encoding is quadratic in variables and cubic in constraints.
So the QBF encoding is promising for instances with many qubits.

In case of S+R synthesis (Table~\ref{tb:stats1}, Experiment 2), adding CNOT restrictions seems to boost the performance of SAT encoding compared to QBF.
Note that the coupling graph is typically planar, with a low out-degree, so the SAT encoding with restrictions becomes quadratic instead of cubic.
For depth optimization, both SAT and QBF techniques take only a few seconds for each slice (see Table~\ref{tb:stats2}).
Since the optimal depths of slices are small, the memory footprint is negligible (close to zero, not shown here).
Only QBF takes around 125 MB memory for the 14-qubit instance \texttt{rcadder6}.

Both W and S+R variants are practical for optimization: most instances are solved within a minute in our benchmark set.
While we cannot apply peephole optimization with W+R, we optimized the individual slices from experiment 2 with W+R.
The W+R encoding performs well. It optimally solves all slices and never takes more than a minute for any slice.

\paragraph*{CP vs SAT and QBF}
In general, fd-ms (CP) is slower than using SAT and QBF solvers (Table~\ref{tb:stats1}).
Overall it results in 5 timeout slices compared to 3 for SAT and QBF.
We also noticed that fd-ms uses more memory (up to 6 GB) compared to the other two.

Note that CP-based solving techniques are orthogonal to SAT and QBF.
For instance, CP results in maximum CNOT depth reduction in Experiment 2 with the instance \texttt{qft4}.
Another advantage of the CP approach is being able to use fast heuristic planners.
Just using any heuristic planner results in heuristic CNOT optimization.
For large circuits (with hundreds of qubits), such an approach is more feasible than the SAT and QBF based approaches.

\section{Related work}
\label{sec:relatedwork}

CNOT synthesis has been studied before, also in the context of qubit permutation and CNOT restrictions.
In this section, we discuss some CNOT synthesis approaches that are close to our approach.

\paragraph*{S and S+R synthesis}
Several techniques are applied for CNOT synthesis such as Gaussian elimination~\cite{beth2001quantum}, Steiner trees~\cite{DBLP:journals/qic/KissingerG20},
rewrite rules~\cite{iwama2002transformation}, and asymptotically optimal algorithms~\cite{DBLP:journals/qic/PatelMH08,DBLP:journals/qic/KissingerG20}.
Optimal CNOT synthesis is mainly considered in a broader context i.e., in the presence of either T gates or RZ gates.
Here instead of synthesis on $n \times n$ matrix, synthesis so-called phase polynomial is applied which also keeps track of phase rotation by T gates.
Synthesis is applied in some polynomial representation using Steiner trees in ~\cite{9914638}, as SAT in~\cite{DBLP:conf/rc/MeuliSM18}, and
as Answer Set Programming (ASP) in~\cite{DBLP:conf/cilc/PiazzaRW23,DBLP:conf/aiqxqia/PiazzaR23}.
Giving CNOT circuits without T or RZ gates as inputs for such encodings results in S and S+R variant encodings.

\paragraph*{W and W+R synthesis}
In~\cite{DBLP:conf/rc/BrugiereBVMA20}, authors proposed heuristic W and W+R variants based on the Syndrome Decoding Problem.
The same authors proposed greedy algorithms for W in DaCSynth, which we compared with in this paper.
Qubit permutations are applied in the TKET compiler~\cite{sivarajah2020t}, but only without CNOT restrictions.
In all variations, allowing qubit permutations results in further reduction in both CNOT count and depth.
To our knowledge, W and W+R variants have not been handled optimally before.

\paragraph*{Beyond CNOT synthesis}
SAT-based Synthesis of Clifford circuits with CNOT, H, and S gates has been proposed with both gate~\cite{Schneider2022ASE} and depth optimization~\cite{Peham2023DepthOptimalSO} in the QMAP tool.
Using CNOT circuits as input in the QMAP tool is similar to our S variant synthesis.

Instead of peephole optimization with circuit slicing, one can apply global CNOT synthesis using so-called holes as in~\cite{DBLP:journals/corr/abs-2308-16496}.
CNOT synthesis is sometimes integrated with Layout Synthesis to achieve further reduction as in the heuristic approaches of~\cite{DBLP:journals/corr/abs-2205-00724,Ding2020ExactSO,wu2023optimization}.

\section{Conclusion}
\label{sec:conclusion}

In this paper, we considered optimal CNOT synthesis with two extensions, qubit permutation and layout restrictions.
To our knowledge, we provide the first optimal CNOT synthesis variants with qubit permutation.
We have encoded variations of optimal CNOT synthesis in Classical Planning, SAT, and QBF.
We handled both CNOT count and CNOT depth metrics for optimization.
By applying peephole optimization, we validated our techniques on standard T-gate optimal benchmarks.
Our results show the effectiveness of qubit permutation on CNOT count and depth reduction.
Finally, we showed further reduction in already optimally mapped benchmarks.

We leave integrated Layout + CNOT Synthesis, including optimal initial mapping,
as a challenge for future work.

\medskip
\medskip
\noindent\textbf{Please cite this paper as:}

\begin{verbatim}
@inproceedings{ShaikvdP2024cnotsynthesis,
author = {Irfansha Shaik and Jaco van de Pol},
title  = {Optimal Layout-Aware CNOT Circuit
          Synthesis with Qubit Permutation},
booktitle = {{ECAI'24}},
address   = {{Santiago de Compostela, Spain}},
publisher = {IOS Press},
year      = {2024}}
\end{verbatim}

\begin{ack}
This research was partially funded by the Innovation Fund Denmark.
The experiments were carried out on the Grendel cluster at the
Centre for Scientific Computing, Aarhus (\url{http://www.cscaa.dk/grendel/}).
We thank the anonymous reviewers of ECAI 2024 for their useful suggestions.
\end{ack}


\bibliography{references}

\end{document}